\documentclass[epj]{webofc}
\usepackage[varg]{txfonts}   % Web of Conferences font
\usepackage{lineno}

\newcommand{\Fermi}{\textit{Fermi}\xspace}
\newcommand{\g}{\ensuremath{\gamma}\xspace}

\begin{document}
%
%\linenumbers
%
\title{\Fermi Bubbles: an Elephant in the Gamma-ray Sky}

\author{Dmitry Malyshev\inst{1, 2}\fnsep\thanks{\email{dmitry.malyshev@fau.de}} 
%\and
}

\institute{
	Erlangen Centre for Astroparticle Physics, Erwin-Rommel-Str. 1, D-91058 Erlangen, Germany, 
\and
         on behalf of the \Fermi LAT collaboration
          }

\abstract{%
  The \Fermi bubbles are one of the most remarkable features in the gamma-ray sky revealed by the \Fermi Large Area Telescope (LAT). 
  The nature of the gamma-ray emission and the origin of the bubbles are still open questions. 
  In this note, we will review some basic features of leptonic and hadronic modes of gamma-ray production.
  At the moment, gamma rays are our best method to study the bubbles, but in order to resolve the origin of the bubbles
  multi-wavelength and multi-messenger observations will be crucial.
}
\maketitle

\section{Introduction}

The \Fermi bubbles were originally discovered as a spherical gamma-ray haze emission \citep{2010ApJ...717..825D}
in a search for a gamma-ray counterpart of the microwave haze detected in the Wilkinson Microwave Anisotropy Probe (WMAP) data \citep{2004ApJ...614..186F}.
With more gamma-ray data, sharp boundaries of the bubbles were resolved at about $55^\circ$ above and below the Galactic center
\citep{2010ApJ...724.1044S}.
The bubbles occupy a solid angle $\sim 1$ sr, which is approximately equal to the area of a square of 3 meters on a side at a 
distance of 3 meters, i.e., about the solid angle of an elephant standing in a room.
The spectrum of the bubbles is $\sim E^{-2}$ between 100 MeV and 100 GeV \citep{2010ApJ...724.1044S, 2014ApJ...793...64A}
with a cutoff or significant softening above 100 GeV \citep{2014ApJ...793...64A}.

Possible origin of the bubbles includes cosmic ray (CR) acceleration by
the supermassive black hole (SMBH) at the center of our Galaxy Sgr A* \citep{2010ApJ...724.1044S},
a period of starburst activity, or an accumulation of CR for a long time from the regular star formation near the Galactic center  (GC)
\cite{2011PhRvL.106j1102C}.
In the latter case, a special arrangement of magnetic fields, e.g., magnetic draping, is necessary to keep the CR from escaping the Galaxy.
Examples of bubble-like structures in other Galaxies include a pair of jets in Centaurus A,
which is also discovered as an extended gamma-ray source \cite{2010Sci...328..725A}, and a star-bust activity in the M82 galaxy.
%As a result, \Fermi LAT gamma-ray data are still the best way to learn about the properties of the bubbles.
To understand the origin of the bubbles, it would be useful to know the gamma-ray production mechanism:
whether the gamma-rays are produced in interactions of hadronic CR with interstellar gas or
by inverse Compton (IC) scattering of high energy electrons with interstellar radiation fields.

\section{Leptonic emission models}

The spectrum of the bubbles can be relatively easily explained with IC gamma rays produced by scattering of interstellar radiation
photons and high energy CR electrons.
The spectrum of the electrons can be represented as a power law with an index $- 2.2$ and an exponential cutoff at 
1.25 GeV \cite{2014ApJ...793...64A}.
There is a tentative association of the \Fermi bubbles emission with the microwave haze
observed in the WMAP \citep{2004ApJ...614..186F} and Planck \citep{2013A&A...554A.139P} data.
For a magnetic field $\approx 8.4 \mu G$ \citep{2014ApJ...793...64A}, the microwave haze can be explained by the same population of electrons 
that produce the gamma-ray emission.
Although the intensity of gamma-ray emission from the bubbles is approximately uniform up to $|b| \approx 55^\circ$ while
the microwave haze intensity decreases significantly above $|b| \approx 35^\circ$,
the difference in the morphology of the gamma-ray and the microwave emission 
can be explained by a decrease in the magnetic field further away from the Galactic plane 
\citep{2012ApJ...750...17D}.

The total gamma-ray luminosity of the \Fermi bubbles $L_{\rm bbl} \approx 4 \times 10^{37} {\rm erg/s}$.
The power of injection of CR electrons should be $\gtrsim L_{\rm bbl}$.
Upscattering of cosmic microwave background (CMB) photons to a few GeV energies requires electron energies $\sim 1$ TeV
($E_\g = 4/3 h\nu \g^2$, where $h\nu \approx 2.4\times 10^{-4}\:\rm eV$ and $\g \sim 2\times 10^6$).
Since the cooling time of a 1 TeV electron is about ${\rm 1\: Myr = 3 \times 10^{13}\: s}$,
the total power contained in electrons above 1 TeV is $\sim 10^{51}$ erg.
Integrating the electron CR spectrum above 1 GeV gives $\approx 10^{52}$ erg \citep{2014ApJ...793...64A}.
Dividing by an approximate volume of the bubbles, one can derive the energy density of the CR electron population.
Around 1 TeV, it turns out to be a factor of 2 to 3 larger than the local energy density of CR electrons \citep{2014ApJ...793...64A},
while around 100 GeV the energy density of electrons inside the bubbles is about the same as the local energy density.

The possibility to explain the \Fermi bubbles together with the microwave haze by the same population of electrons is very appealing,
one difficulty of the leptonic interpretation of the gamma-ray emission is the necessity to have electrons of $~ 1$ TeV energies
around 10 kpc away from the Galactic plane (GP).
If the electrons were produced near the GC, then, with the cooling time of $\lesssim 1$ Myr, the velocity of their transport to high latitudes should
be at least ${\rm 10\: kpc / 1\: Myr \approx 10^4\: km\: s^{-1}}$. 
Since this velocity is much larger than the speed of sound in the plasma around the GP,
one would expect to see a shock front at the boundary of the bubbles, but no such shock front has been observed.
Moreover, the observed velocities of gas outflow in the direction of the bubbles
are $\lesssim 10^3\:\rm km\: s^{-1}$ \cite{2015ApJ...799L...7F} .
A possible solution is that the electrons are (re)accelerated via the 2nd order Fermi acceleration mechanism
by the sound waves left behind a shock front that may have existed as the bubbles were forming \cite{2011PhRvL.107i1101M}
or from a series of shocks expanding from the Sgr A* \cite{2011ApJ...731L..17C}.

Overall, leptonic model offers natural explanations for the gamma-ray spectrum of the bubbles and for the 
possibly associated microwave emission.
The uniform spectrum of the bubbles as a function of latitude can be explained with re-acceleration of electrons
by an ensemble of shock waves inside the bubbles.

\section{Hadronic emission models}

The second possible explanation of the gamma-ray emission from the \Fermi bubbles is the
interactions of high energy CR hadrons with the interstellar gas.
The density of gas above a few hundred parsecs is at least a 100 times smaller than the density of gas in the GP,
the energy density of CR required to explain the emission from the bubbles (which is about 10 - 30 times smaller than the 
gamma-ray emission from hadronic interactions in the GP above 10 GeV)
is 3 - 10 times larger than the CR density in the GP above 100 GeV \cite{2014ApJ...793...64A}.
The spectrum of CR protons that can explain the gamma-ray spectrum of the bubbles is $\sim E^{-2}$ with a cutoff around 10 TeV.
There is a slight tension of the primary gamma-ray spectrum with the observed spectrum of the bubbles around 100 MeV
due to the pion cutoff.
This tension can be easily resolved if one takes into account secondary IC emission from the electrons and positrons produced in the hadronic
interactions together with the primary gamma rays.
However, the spectrum of the secondary leptons is too soft to explain the microwave haze \cite{2014ApJ...793...64A}. Since IC gamma rays above 100 MeV are mostly produced by leptons with $E \gtrsim 100$ GeV up-scattering CMB photons, while the microwave haze is produced by electrons with energies $\lesssim 30$ GeV for a magnetic field $\gtrsim 5\: \rm \mu G$, one needs a population of electrons below 100 GeV with a hard spectrum to explain the microwave haze in addition to the secondary leptons in hadronic interactions.

Integrating the energy density of CR hadrons over the volume of the bubbles, 
one gets the total energy contained in CR protons above 1 GeV of $3 \times 10^{55}$ erg (for the gas density of $0.01\: \rm cm^{-3}$).
Assuming the scattering cross section of 30 mb, the characteristic interaction time is $3 \times 10^9$ yr,
i.e., unless there is a mechanism to keep the protons inside the bubbles for several billions of years,
the majority of protons will escape the bubbles without interacting.
If there is a magnetic draping on the boundary of the bubbles, it can reflect the protons back into the volume of the bubbles
and prevent the escape.
In this case the energy density of the CR can be obtained by accumulating the output in CR from about $3\times 10^5$ supernovae (SN)
assuming that a characteristic kinetic energy of a SN shell is $10^{51}$ erg and efficiency of CR acceleration is 10\%
(see, e.g., discussion in \citep{2011PhRvL.106j1102C}).
This can be easily accumulated over $\sim$ billion of years assuming a SN rate in the inner part of the Galaxy $\sim 1\:\rm kyr^{-1}$.

If we divide, the total energy in CR to the volume of the bubbles (e.g., assuming that the bubbles consist of two spheres with 5 kpc radii),
then we get the energy density in CR of $\sim 1\rm\: eV\: cm^{-3}$.
This energy density can be compared to the kinetic energy density of plasma in the halo around the GP
$\rho \approx 0.0035\:\rm cm^{-3}$ and $kT \sim 340\;\rm eV$ \cite{1997ApJ...485..125S}, 
which gives $W_{\rm plasma} \sim 1.5\rm\: eV\: cm^{-3}$.
The conclusion is that the kinetic energy of the CR is sufficient to push away the plasma and create a cavity
that may be observed in X rays.
The energy of CR electrons in leptonic scenario $\sim 10^{52}$ erg is at least three orders of magnitude smaller than the  energy density of CR protons. As a result, one should not expect to see such cavity in leptonic models.

Although no clear cavity has been observed in ROSAT data, there are preliminary signs of emission measure change across
the boundary of the bubbles with pointed observations by Suzaku \cite{2013ApJ...779...57K}.
The presence of the cavity may be confirmed or excluded by the future more sensitive all-sky X-ray survey with eROSITA.

\section{Models of \Fermi bubbles creation}

The two most popular models of creation of the bubbles are an active galactic nucleus (AGN)-like
activity of the SMBH Sgr A* at the center of our Galaxy and a starburst near the GC.
Numerical simulations show that the bubbles can be inflated by two jets emitted from the SMBH
\cite{2012ApJ...756..181G, 2012ApJ...756..182G, 2012ApJ...761..185Y},
as well as by a spherical outflow, shaped into the bubbles by molecular clouds around the GC \cite{2011MNRAS.415L..21Z}.

Since gamma-ray data have no jet-like structure associated with the bubbles,
one has to rely on external data to distinguish AGN jets or an outflow from a starburst-driven wind 
as the main mechanism that has created the bubbles.
Observations of external galaxies suggest \cite{2010ApJ...711..818S} that AGN winds come 
together with significant photo-ionization, while starburst winds are associated with shock-ionization.
Although the ultraviolet (UV) radiation during star formation has comparable luminosity to UV radiation from accretion onto the black hole,
the UV radiation from young stars is largely absorbed by molecular clouds, where the star formation typically happens,
in addition, there is a delay of several Myr or more between the star formation and SN explosions, which drive the wind 
\cite{2010ApJ...711..818S}.
Unfortunately no ionization has been detected near the \Fermi bubbles that could help to distinguish the two scenarios.
There is, however, an enhanced H$\alpha$ emission in the Magellanic Stream towards the South Pole in Galactic coordinates,
which is consistent with UV radiation produced in an AGN-like activity of Sgr A* around $0.5 - 3$ Myr ago \cite{2013ApJ...778...58B}.
Although such level of ionization in the Magellanic Stream is improbable in the starburst scenario \cite{2013ApJ...778...58B},
it does not exclude the possibility that a starburst, e.g., star formation related to $\sim 6$ Myr population of stars near the GC
\cite{2006ApJ...643.1011P}, produced the bubbles.

\section{Conclusions}

Although the \Fermi bubbles are some of the brightest and most significant features in the gamma-ray sky, 
their origin and even the nature of the gamma-ray emission remains unresolved.
Both leptonic IC scattering and hadronic production of gamma rays are viable options,
while the formation of the bubbles can be either due to an AGN-like activity of Sgr A* or a starburst event near the GC.

One of the signatures of leptonic origin of the gamma-ray emission is the synchrotron emission from the electrons.
There is a tentative association of the microwave haze emission and the \Fermi bubbles which can be explained by the same population
of electrons.
A discovery of polarized emission with features correlated with the \Fermi bubbles would strengthen this hypothesis.

For the hadronic model, the required CR energy density is comparable to the energy density of the halo plasma.
As a result, the presence of the CR may create a cavity in the plasma, that can be detected with the future X-ray observations,
most notably, eROSITA.
There is also a possibility to detect an associated neutrino signal with new more sensitive neutrino detectors,
such as PINGU in IceCube and KM3NeT.

Developing a model of the \Fermi bubbles near the GC will play a significant role in disentangling their origin.
If the bubbles were produced by a Sgr A* activity, then they should have a narrow base centered at the GC,
while a starburst activity may result in a broader base of the bubbles not necessarily centered on the GC.
Current and future Cherenkov telescopes can help to study the morphology of the \Fermi bubbles near the GC
at energies $\gtrsim 100$ GeV, where the \Fermi LAT loses sensitivity.

\vspace{1mm}
The \textit{Fermi}-LAT Collaboration acknowledges support
from NASA and DOE (United States), CEA/Irfu, IN2P3/CNRS, and CNES (France), ASI, INFN, and INAF (Italy), MEXT, KEK, and JAXA (Japan), and the K.A.~Wallenberg Foundation, the Swedish Research Council, and the National Space Board (Sweden). 
This work was partially supported by NASA grant NNH13ZDA001N.
\vspace{-1mm}

%
% BibTeX or Biber users please use (the style is already called in the class, ensure that the "woc.bst" style is in your local directory)
\bibliography{malyshev_bubbles_ricap16_papers.bib}
%
%%%%%%%%% Non-BibTeX users please use
%%%%%%%%%
%%%%%%%%\begin{thebibliography}{}
%%%%%%%%%
%%%%%%%%% and use \bibitem to create references.
%%%%%%%%%
%%%%%%%%\bibitem{RefJ}
%%%%%%%%% Format for Journal Reference
%%%%%%%%Journal Author, Journal \textbf{Volume}, page numbers (year)
%%%%%%%%% Format for books
%%%%%%%%\bibitem{RefB}
%%%%%%%%Book Author, \textit{Book title} (Publisher, place, year) page numbers
%%%%%%%%% etc
%%%%%%%%\end{thebibliography}

\end{document}